\documentclass[10pt, a4paper]{article}
\usepackage{epsfig}
\usepackage{graphicx}               
\usepackage{amsmath}                
\usepackage{multirow}               
\usepackage{fancyhdr}
\usepackage{here}

\begin{document}

\pagestyle{fancy}

\title{\LARGE\bf Performance study of SKIROC2 and SKIROC2A with BGA testboard}
\author{
  {\large I.~Sekiya$^1$\thanks{Presenter. Talk presented at the International Workshop on Future Linear Colliders (LCWS16), Morioka, Iwate, 5-9 December 2016. i\_sekiya@epp.phys.kyushu-u.ac.jp}, H.~Hirai$^1$, T.~Suehara$^1$\thanks{suehara@phys.kyushu-u.ac.jp}, T.~Yoshioka$^2$, K.~Kawagoe$^1$} \\ \\
  $^1$Department of Physics, Faculty of Science, Kyushu University \\
  $^2$Research Center for Advanced Particle Physics, Kyushu University \\
  744 Motooka, Nishi-ku, Fukuoka, 819-0395 Japan
 }
\date{}

\maketitle
\thispagestyle{fancy}
\setcounter{page}{1} 

\begin{abstract}
SKIROC2 is an ASIC to readout the silicon pad detectors for the electromagnetic calorimeter in the International Linear Collider.
Characteristics of SKIROC2 and the new version of SKIROC2A, packaged with BGA, are measured with testboards and charge injection. The results on the signal-to-noise ratio of both trigger and ADC output, threshold tuning capability and timing resolution are presented.
\end{abstract}

\section{Introduction}

 The International Linear Collider (ILC) is a next-generation electron-positron linear collider, which is planned to be constructed in Japan. The main targets of the ILC include precise measurements of
 the Higgs boson and the top quark, and searches for particles of new physics.
The center of mass energy of the ILC is 250GeV to 500 GeV with 31 km length, with possible upgrade to 1 TeV.

\begin{figure}[htbp]
\begin{center}
\includegraphics[width=80mm]{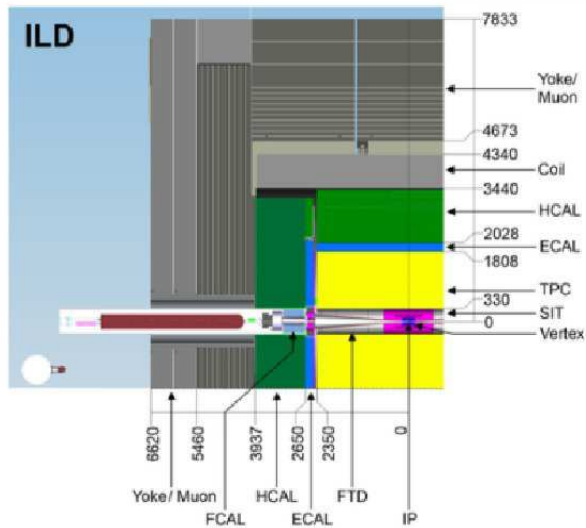}
\caption{Cross section of the ILD detector.}
\label{fig:ILD}
\end{center}
\end{figure}

 The International Large Detector (ILD) \cite{ILD} is one of two detector concepts of ILC (Fig.~\ref{fig:ILD}). The ILD consists of a vertex detector, silicon trackers and time projection chamber for tracking charged particles, electromagnetic and hadron calorimeters (ECAL and HCAL), and a superconducting solenoid with a muon tracker in its return yoke. The key concept of the ILD is ``Particle flow" \cite{pfa}, which is to obtain supreme jet energy resolution by separating particles inside the jets and assign each track to a calorimeter cluster one by one. To maximize the particle separation, high granularity at both ECAL and HCAL is critical.

The ILD ECAL is a sampling calorimeter with tungsten absorber and highly segmented readout layers. Silicon-Tungsten ECAL (SiW-ECAL), which employs silicon pad sensors for the readout is the reference option for the ILD ECAL. In the design of SiW-ECAL 5 $\times$ 5 mm cells are implemented in the silicon pads, which results in around 100 million channels in total of 30 layers of ILD ECAL. The readout ASIC should be embedded between absorber and detector layers. The ASIC and related readout electronics are the major development issues for the ILD ECAL, especially for its dense and efficient readout and the strict limit of heat discipation.

\section{SKIROC2 and testboards}

\begin{figure}[htbp]
\begin{center}
\includegraphics[width=130mm]{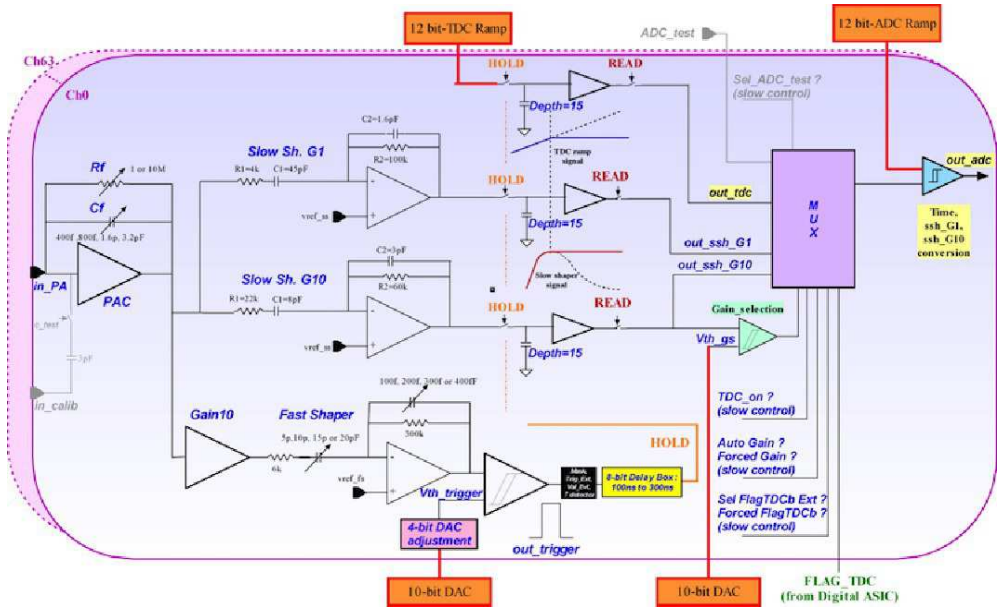}
\caption{A schematic diagram of the analog part of SKIROC2.}
\label{fig:skiroc2}
\end{center}
\end{figure}

 SKIROC2 (Silicon Kalorimeter Integrated ReadOut Chip 2) \cite{skiroc2} is an Application-Specific Integrated Circuit (ASIC) for the ILD SiW-ECAL, developed by IN2P3/Omega group. A schematic diagram of the analog part of SKIROC2 is shown in Fig.~\ref{fig:skiroc2}. It has 64 input channels, with wide acceptable signal strength of 0.5-2500 MIP (Minimum Ionizing Particle) equivalent charges. The signal from detectors or the test pulse input passes through the preamplifier with variable gain set by the feedback capacitance $C_F$. The stability of the preamplifier is controlled by a variable compensation capacitance. The output of the preamplifier is sent to a fast shaper and two slow shapers with different gain. The fast shaper output is used for self-triggering function. The trigger signal is used to hold the voltages at two shaper outputs and an additional scanned voltage source for timing measurements, which are sent to 15-depth analog memory cells. The trigger threshold can be adjusted by a global digital voltage reference, with a fine-tuning capabilty for the threshold of individual channels. The charges in the memory cells are readout after the acquisition by a 12-bit Analog-to-Digital Converter (ADC) and a multiplexer, with a bunch ID tagged with 5 MHz slow clock.
 
 A special function of SKIROC2 is called ``power pulsing" which is a power control function synchronized and optimized for the ILC bunch structure. In the ILC, 1312-2625 bunches (depending on the run condition) of electrons, called ``a train", come to the collisions within around 1 ms duration with 2-5 MHz collision frequency. The operation frequency of the ILC is 5 Hz, so there is a 199 ms pause between the trains. The power pulsing is to switch off unnecessary part of SKIROC2 on each stage of readout procedure to reduce the heat dissipation. The power supply of SKIROC2 is separated into four parts, analog, DAC, ADC and digital. For example, the current of the analog part is only supplied during the acquisition, and the current is shut up during readout phase.

 SKIROC2A is a new version of SKIROC2, which came to be available since last year. Modifications from SKIROC2 includes wider dynamic range of the trigger threshold adjustment of individual channels, reducing fake triggers occurred at the specific timing in SKIROC2, improving delay cells used for trigger delay, automatic gain selection function, etc. The main functions of the two versions of SKIROC2 are examined and signal-to-noise (S/N) ratio is compared between the two ASICs in this study.

The SKIROC2 is available on the Ball Grid Array (BGA) package, as well as Quad Flat Package (QFP).
The BGA package have grid contacts (400 contacts in SKIROC2) on the bottom, with a smaller footprint than the QFP package, and it is more robust than directly bonding a naked die to a Printed Circuit Board (PCB).

To evaluate the functions and performance of SKIROC2 and SKIROC2A, we designed a testboard \cite{Hiraitalk, Hirairef} dedicated for BGA SKIROC2/SKIROC2A, based on a design of similar boards for QFP SKIROC2 by Omega group. Updates from the Omega board includes a connector for silicon pads and a interface to ILD SiW-ECAL electronics. The testboard is equipped with an Altera FPGA. Control of the FPGA can be done via a USB interface and a Labview software, while the data aquisition is done with a C++ software via the USB interface, which outputs data format compatible with the ILD SiW-ECAL data aquisition system.

\begin{figure}[htbp]
\begin{minipage}{0.5\hsize}
\begin{center}
\includegraphics[width=60mm]{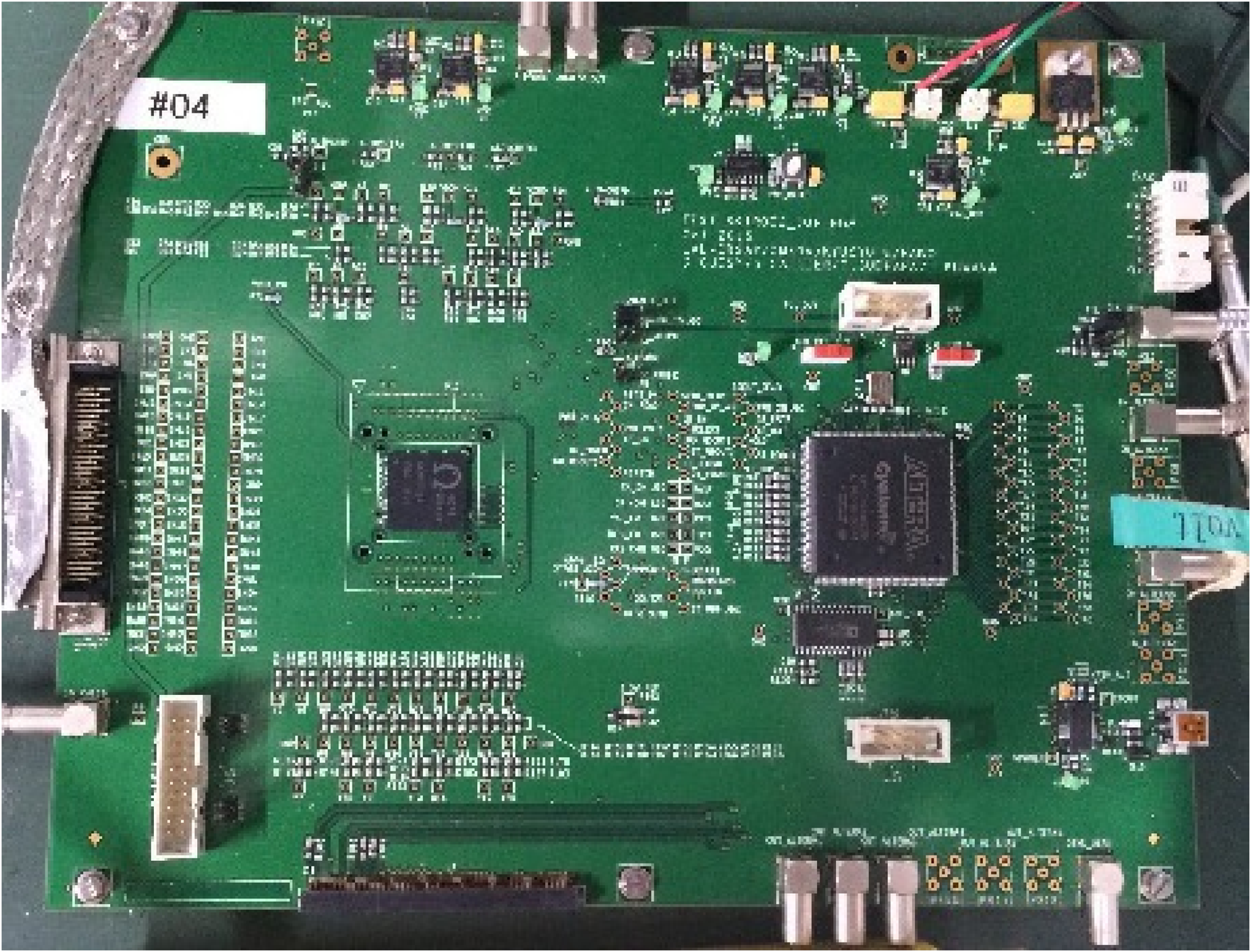}
\end{center}
\end{minipage}
\begin{minipage}{0.5\hsize}
\begin{center}
\includegraphics[width=60mm]{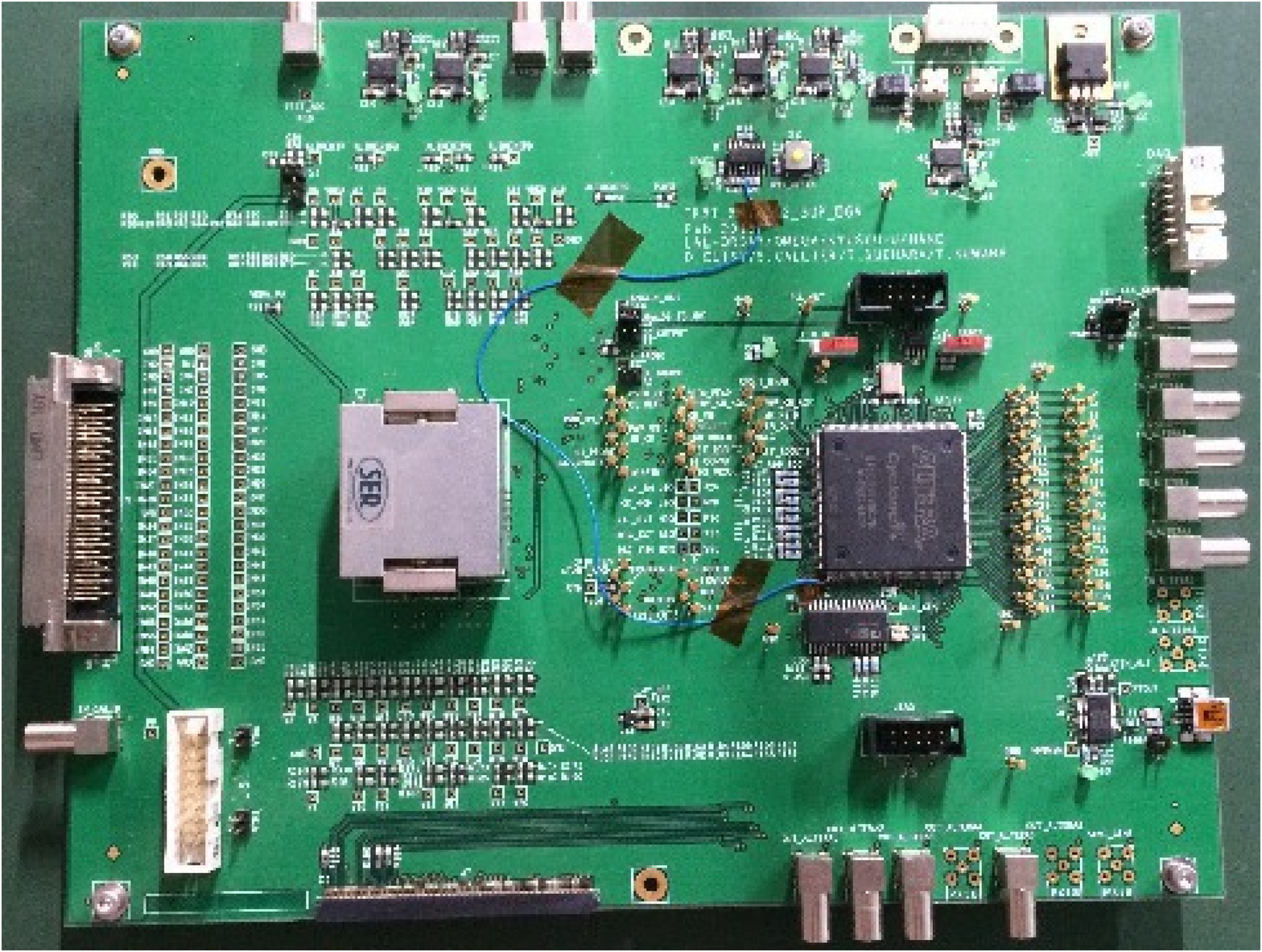}
\end{center}

\end{minipage}
\label{fig:testboard}
\caption{Testboard 1 (left) and Testboard 2 (right). At Testboard 1, the ASIC is mounted with a socket. At Testboard 2, SKIROC2A chip is directly soldered on the board.}
\end{figure}

We fabricated two types of testboards, shown in Fig.~\ref{fig:testboard} for the SKIROC2/SKIROC2A in the BGA package. The one is equipped with a BGA400 socket, called ``Testboard 1", and the other is with a soldered SKIROC2A, called ``Testboard 2". The SKIROC2 and SKIROC2A can be easily replaced in the Testboard 1, which gives easier comparison of two ASICs. In contrast, the Testboard 2 has no function of replacement, whereas the noise level is expected to be better in the Testboard 2.

\section{Measurements on trigger}

\begin{figure}[htbp]
\begin{center}
\includegraphics[width=80mm]{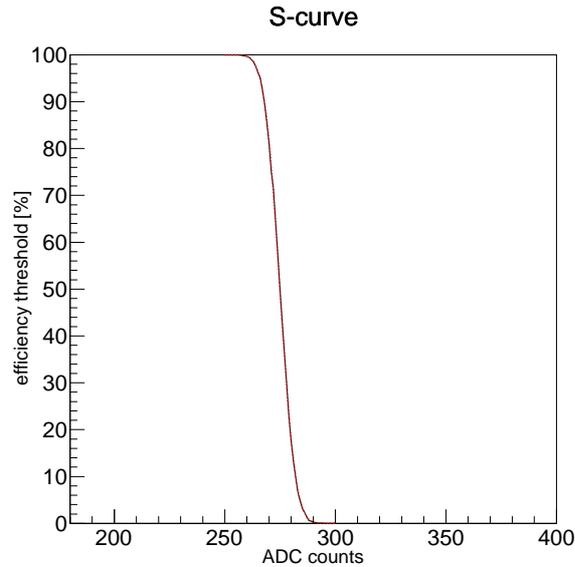}
\caption{A sample of the s-curve. The horizontal axis is the trigger threshold, and the vertical axis is the trigger efficiency.}
\label{fig:scurve}
\end{center}
\end{figure}

The fast shaper output is used for triggering. The S/N ratio of the triggering is obtained via the ``S-curve" scan, shown in Fig.~\ref{fig:scurve}. A fixed amount of charge is introduced from the test pulse input. The trigger efficiency to the charge is obtained with varying the trigger threshold, to obtain the curve shown in the plot. The curve is fitted by a complementary error function, whose center value stands for the threshold on the charge, and the sigma stands for the noise power. We obtain the S-curve with 1 and 2 MIP equivalent charges, and the 1 MIP gain is obtained by the difference of the fitted center value with the two configuration. The S/N ratio can be obtained by the noise divided by the gain.
The performance on S/N ratio of triggering is compared with various gain and between SKIROC2 and SKIROC2A. We also measured the dynamic range of individual trigger threshold adjustment, which is significantly improved in SKIROC2A. The all measurements on this section are done with the power pulsing function disabled.

\begin{figure}[htbp]
\begin{tabular}{c}

\begin{minipage}{0.33\hsize}
\begin{center}
\includegraphics[width=50mm]{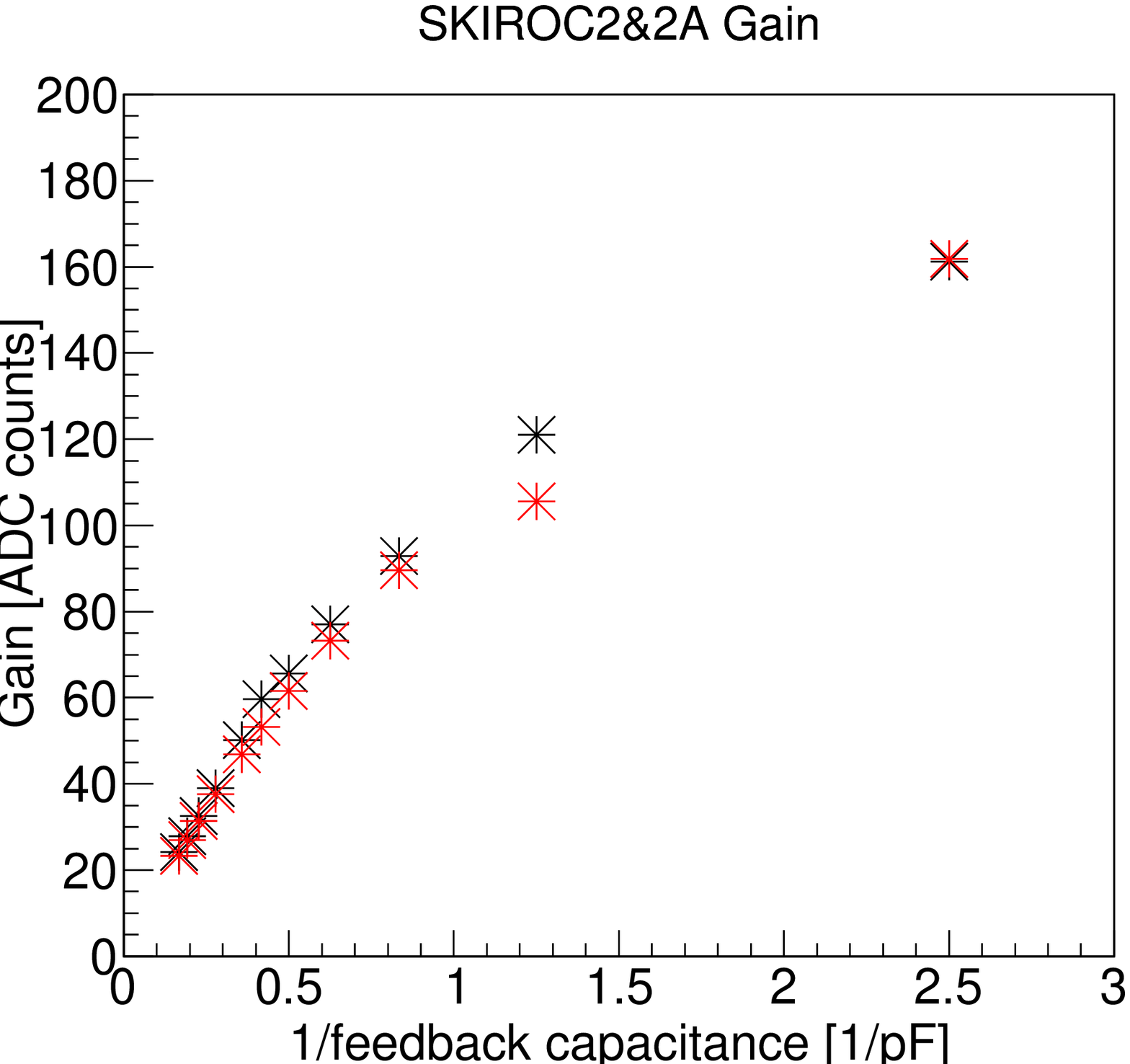}
\end{center}
\end{minipage}

\begin{minipage}{0.33\hsize}
\begin{center}
\includegraphics[width=50mm]{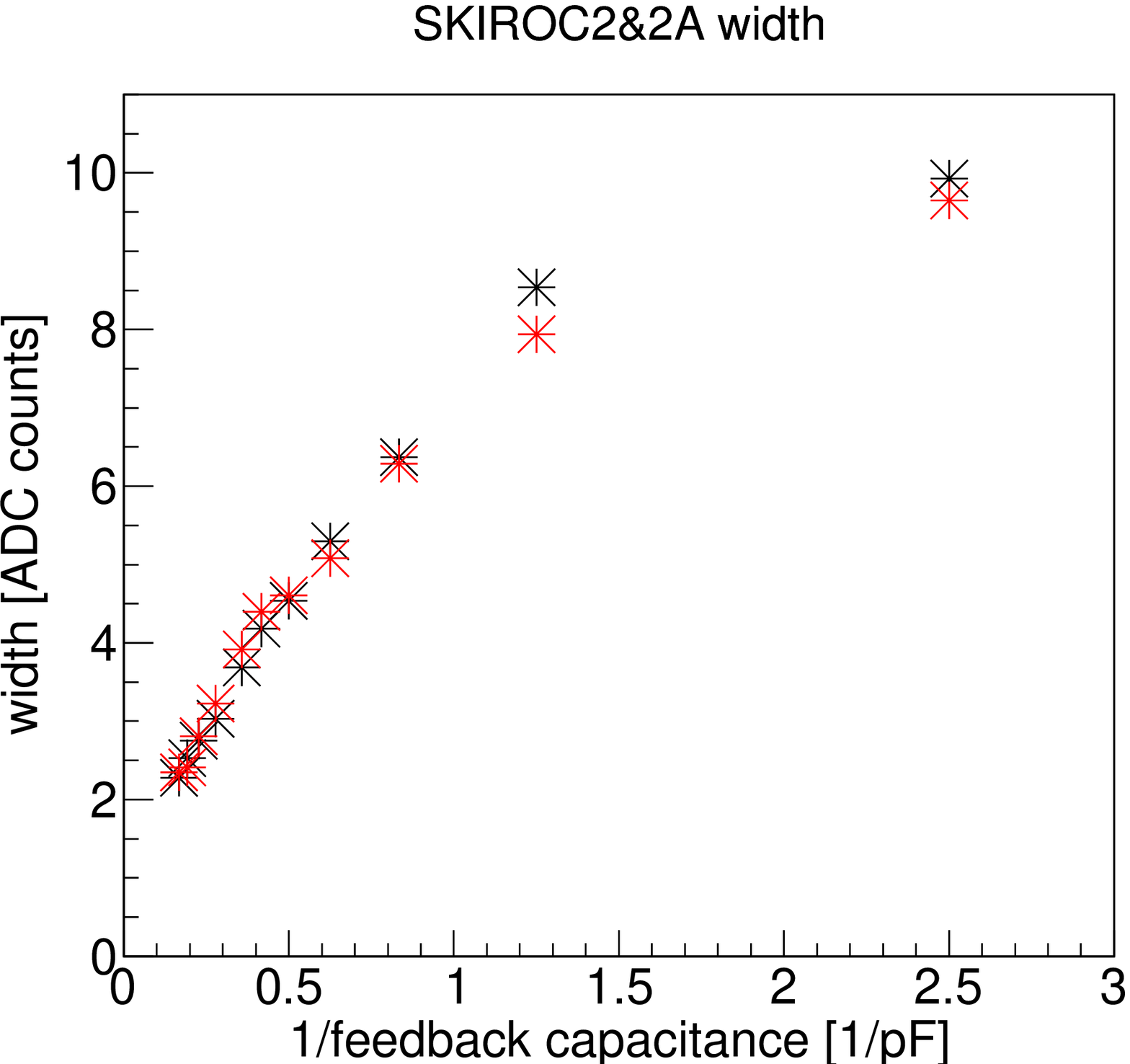}
\end{center}
\end{minipage}

\begin{minipage}{0.33\hsize}
\begin{center}
\includegraphics[width=50mm]{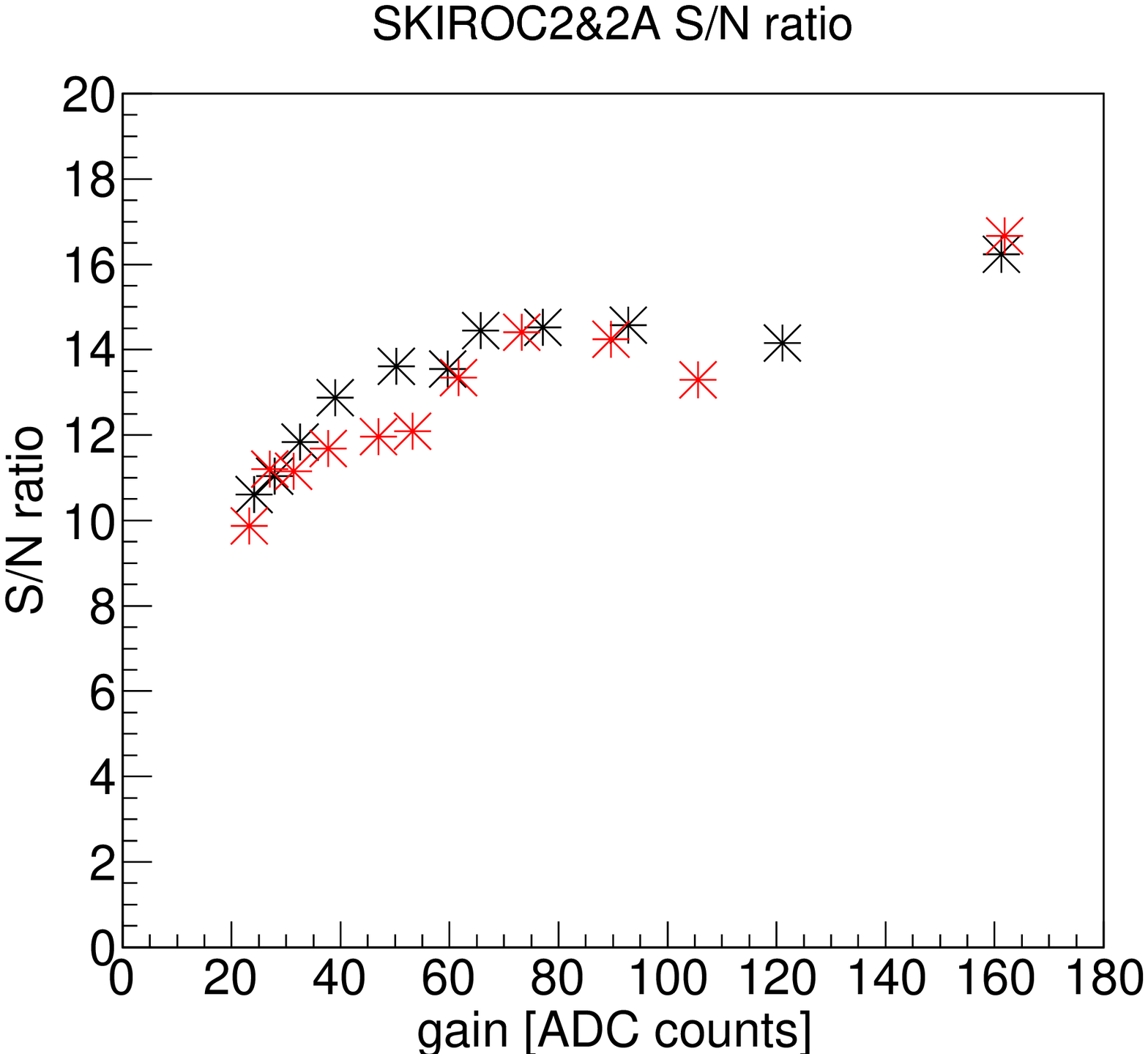}
\end{center}
\end{minipage}

\end{tabular}

\label{fig:fFS}
\caption{The gain, the width and the S/N ratio of SKIROC2 and SKIROC2A with the inversed feedback capacitance on the horizontal axis. SKIROC2 is shown in the black points and SKIROC2A is shown in the red points.}

\end{figure}

Figure \ref{fig:fFS} shows the dependence of the gain, the noise width and the S/N ratio, with varying the feedback capacitance. The compensation capacitance is fixed to 6.0 pF. Channel 10 of SKIROC2 and SKIROC2A are examined with Testboard 1.
We see a almost linear dependence of the gain on the inverse of the feedback capacitance.
The dependence of S/N ratio on the gain is moderate, with slight preference of higher gain to obtain better S/N ratio. No significant difference is seen between SKIROC2 and SKIROC2A.

\begin{figure}[htbp]
\begin{tabular}{c}
\begin{minipage}{0.33\hsize}
\begin{center}
\includegraphics[width=50mm]{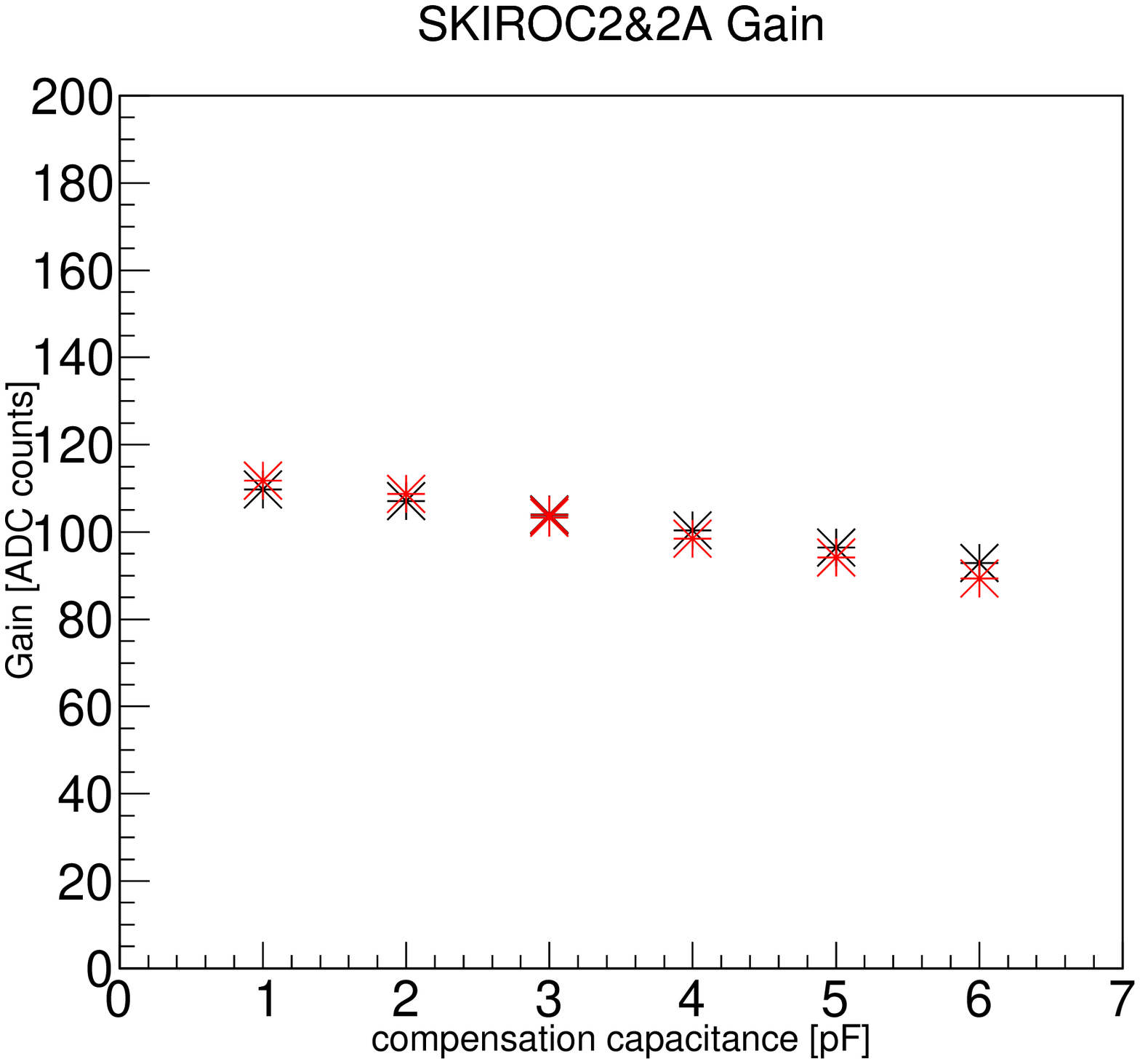}
\end{center}
\end{minipage}

\begin{minipage}{0.33\hsize}
\begin{center}
\includegraphics[width=50mm]{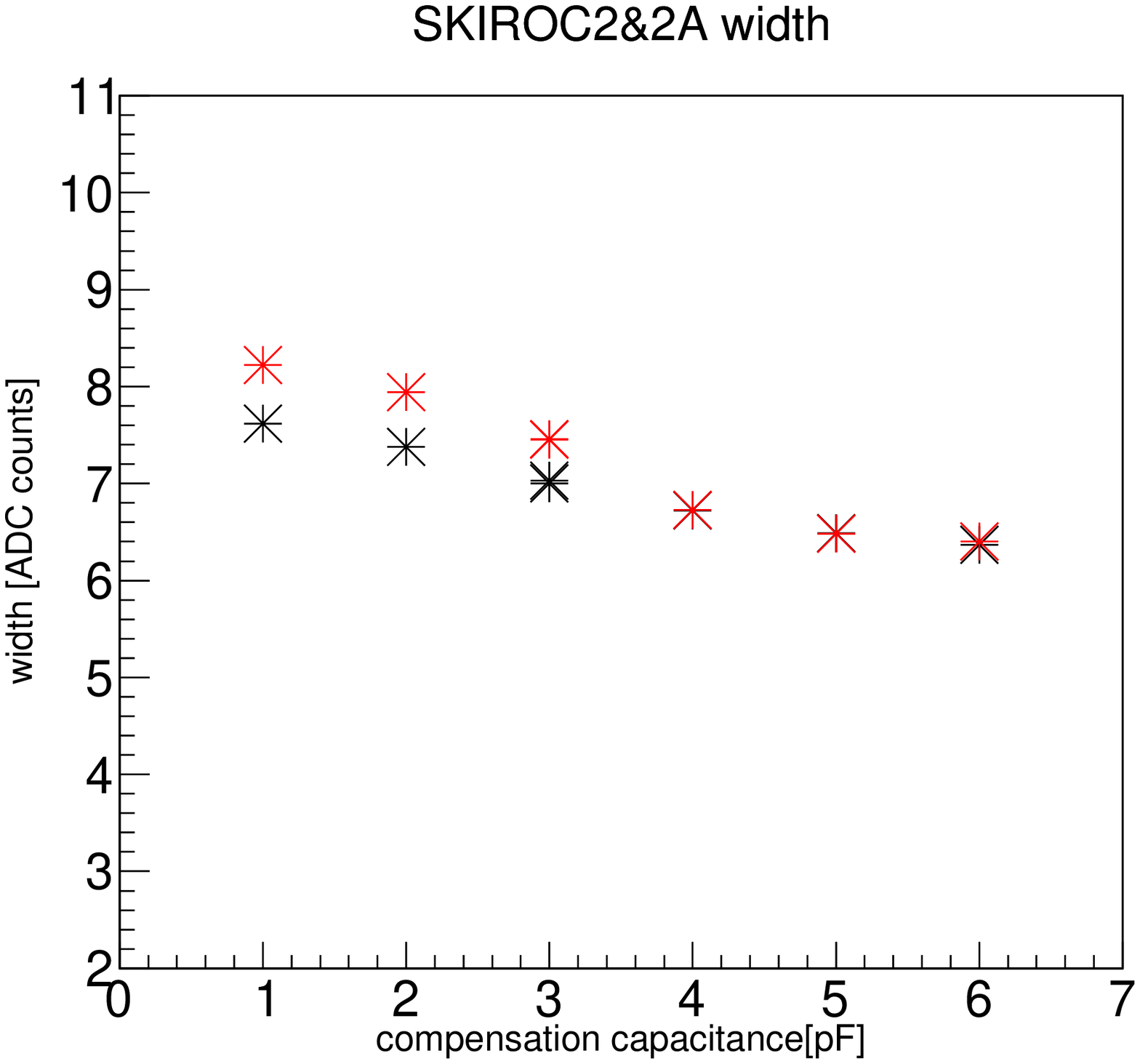}
\end{center}
\end{minipage}

\begin{minipage}{0.33\hsize}
\begin{center}
\includegraphics[width=50mm]{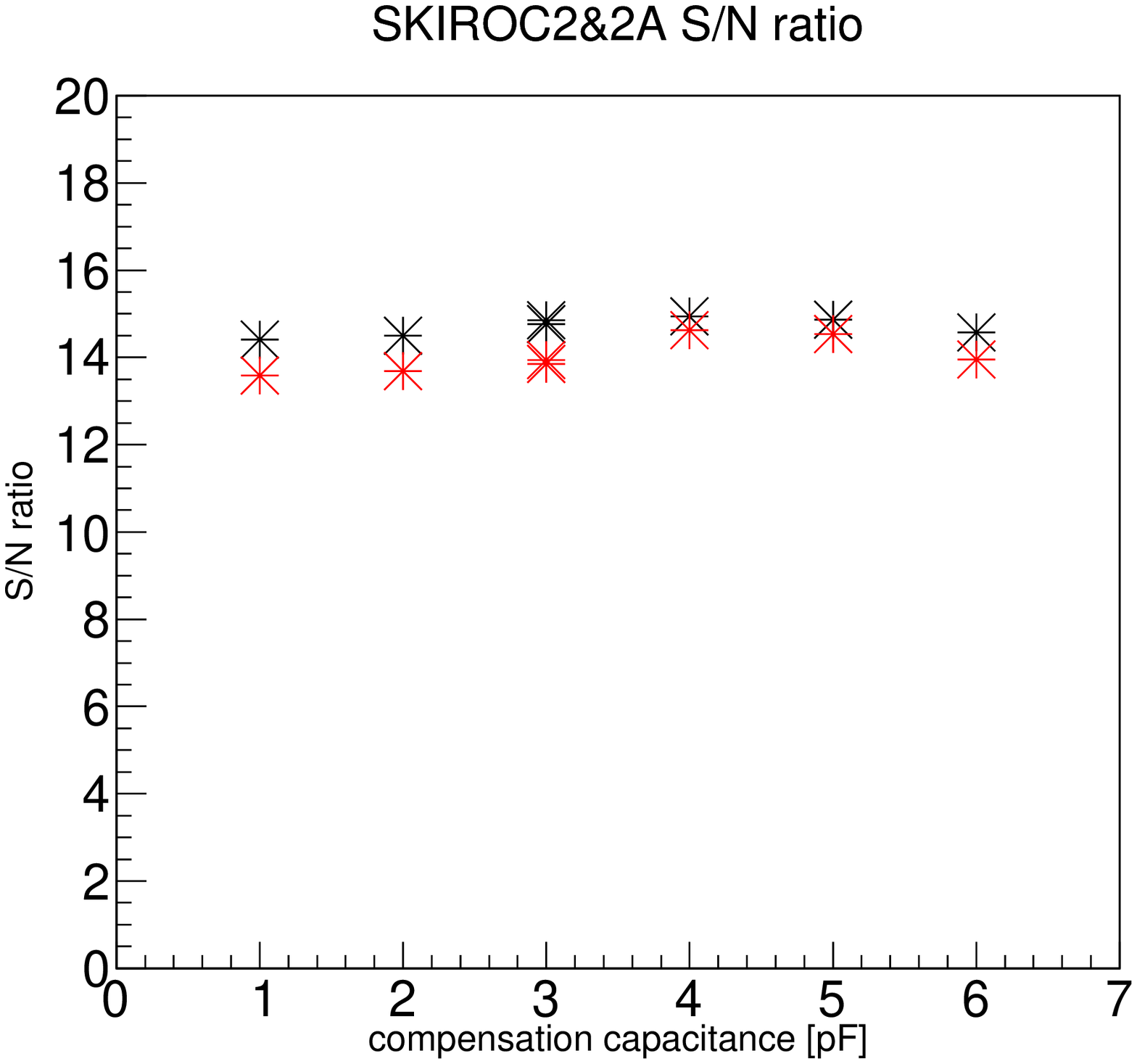}
\end{center}

\end{minipage}

\end{tabular}
\label{fig:cFS}
\caption{Gain, width and S/N ratio with changing the compensation capacitance. The black point is shown as SKIROC2 and the red one is shown as SKIROC2A.}

\end{figure}

 Figure \ref{fig:cFS} shows the dependence on the compensation capacitance, with a fixed feedback capacitance of 1.2 pF. The channel 10 of the Testboard 1 is examined again. As shown in the plot, we see slightly higher gain with the lower compensation capacitance, but the S/N ratio has no big difference on the compensation capacitance. Response of SKIROC2 and SKIROC2A is also very similar.

\begin{figure}[htbp]
\begin{center}
\includegraphics[width=60mm]{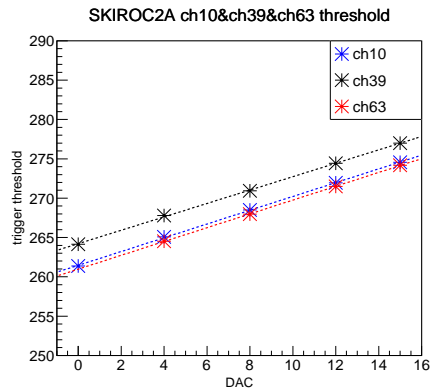}
\caption{Dependence on the global trigger threshold (vertical axis) on the 
trigger adjustment DAC of individual channels (horizontal axis),
examined with channels 10, 39 and 63.
The slopes of the fit lines are around 0.87 for the all three shown channels.
}
\label{fig:chbych}
\end{center}
\end{figure}

\begin{figure}[htbp]
\begin{center}
\includegraphics[width=80mm]{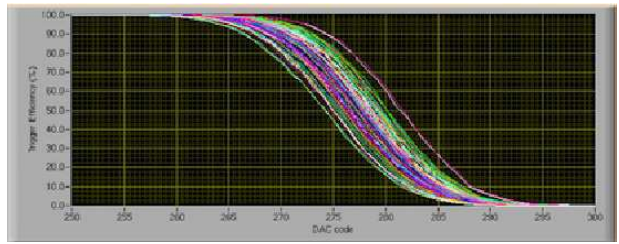}
\caption{S-curve scan for all 64 channels.
The variation on the DAC count with 50\% trigger efficiency among the channels
is around 7 DAC counts, which are within the dynamic range of individual threshold adjustment.}
\end{center}
\label{fig:63chthreshold}

\end{figure}

The trigger threshold is set by two DAC (Digital-to-Analog Conversion) settings, a global threshold
with a 10-bit DAC and channel-by-channel adjustment with 4-bit DACs.
Adjustment on individual channels is tested with SKIROC2A.
The dynamic range is expected to be improved from SKIROC2, which has accidentaly
tiny dynamic range equivalent to around 1 DAC count of the global threshold setting.
Figure \ref{fig:chbych} shows the trigger threshold with 1 MIP test pulses in 
the channels 10, 39 and 63, with varying the individual trigger threshold from 0 to 15 DAC counts.
The observed thresholds are almost linear to the DAC counts, with a dynamic range of around 12.25.
Since the channel to channel fluctuation of the threshold is around 7 global DAC counts as shown in
Fig.~\ref{fig:63chthreshold}, the dynamic range is reasonably wide with SKIROC2A.

\section{S/N ratio on slow shaper}

\begin{figure}[htbp]
\begin{minipage}{0.5\hsize}
\begin{center}
\includegraphics[width=60mm]{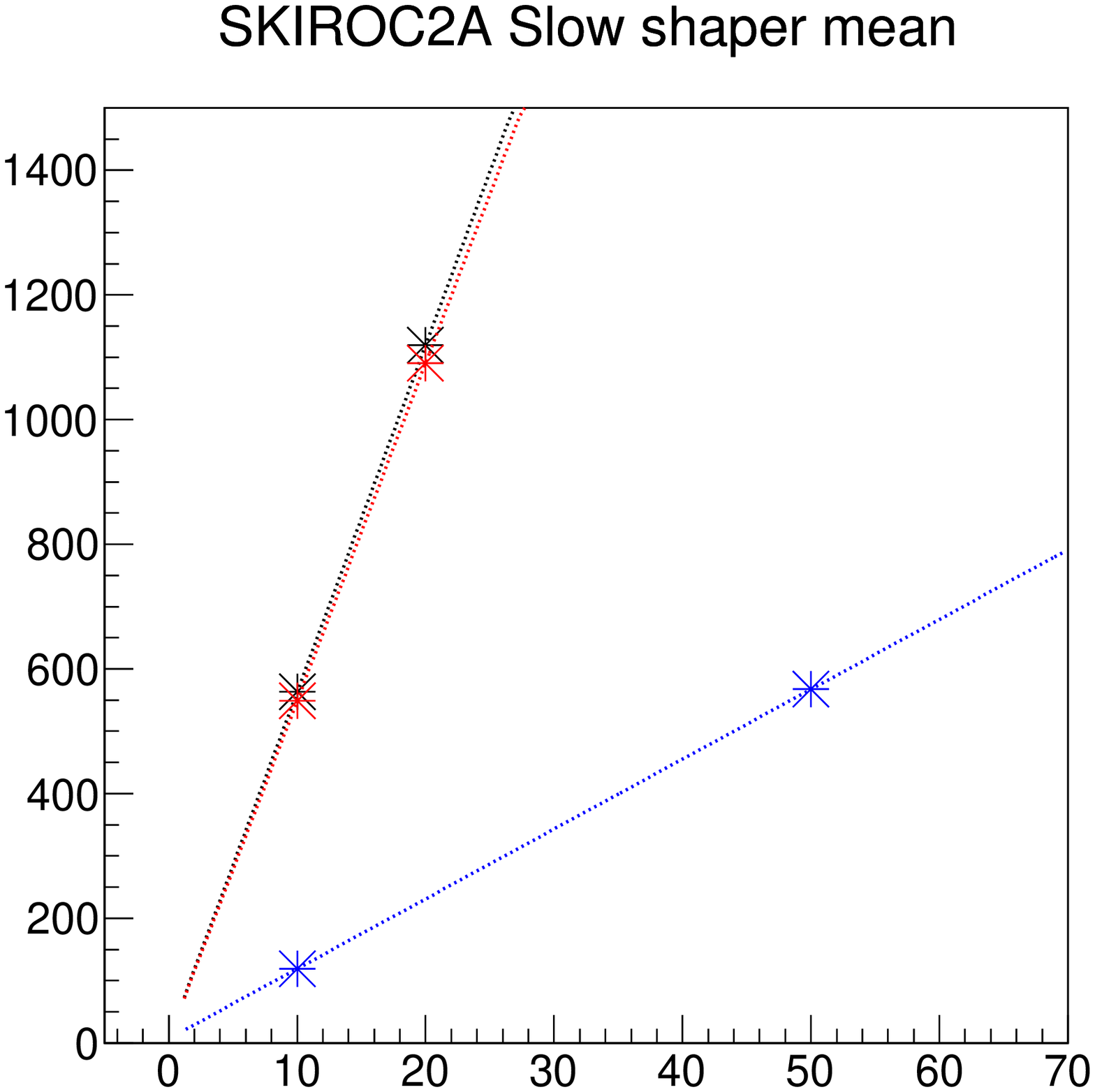}
\end{center}
\end{minipage}
\begin{minipage}{0.5\hsize}
\begin{center}
\includegraphics[width=60mm]{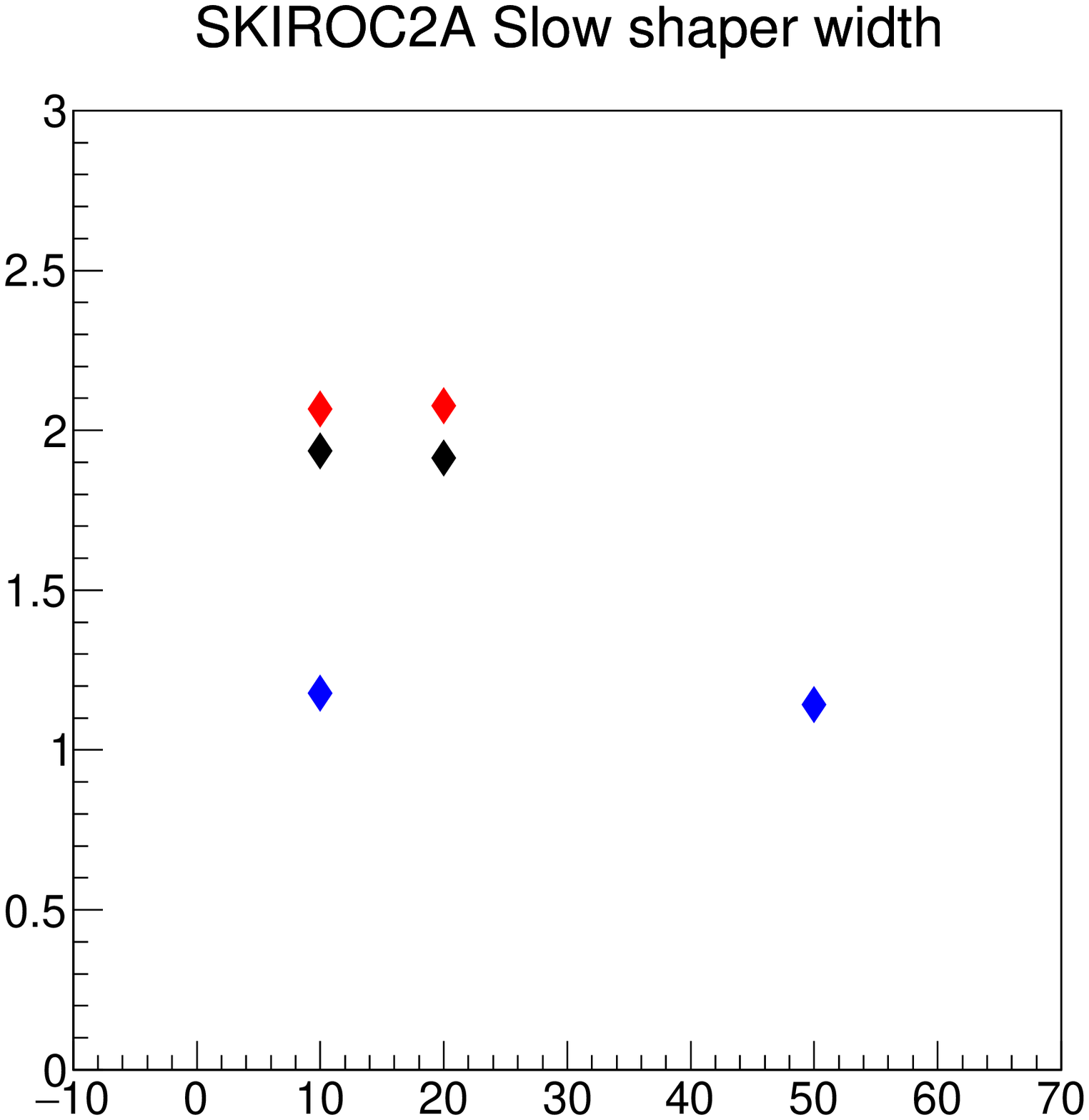}
\end{center}

\end{minipage}
\label{fig:SS}
\caption{ADC mean of signal (left) and pedestal width (right) with charge injection.
The black points show the results with 1.2 pF feedback capacitance and no power pulsing, the red points show the results with power pulsing. The blue points show the results with 6.0 pF feedback capacitance and no power pulsing.}
\end{figure}

The S/N ratio of the slow shaper output on SKIROC2A is examined with Testboard 2.
Figure \ref{fig:SS} shows the ADC mean and the pedestal width with charge injection
for the feedback capacitance of 1.2 pF and 6.0 pF.
The performance with and without power pulsing is also compared in the plot.
For the power pulsing, we only switch the analog part of the power supply, which is
mainly related to the S/N ratio. Delay of 3 ms with respect to the power switching
is added before activating aquisition clock to stabilize the outputs of amplifiers,
which is longer than the operation with SiW-ECAL data acquisition.
This is due to more capacitance added on the testboard
to stabilize the power supplies.

With the 1.2 pF feedback capacitance, the slope is 55 counts per MIP without power pulsing
and 54 counts per MIP with power pulsing. The higher feedback capacitance of 6.0 pF gives
the slope of 11 count per MIP, almost 5 times lower gain as expected.
For the S/N ratio, 1.2 pF feedback capacitance gives S/N ratio of around 29 without power pulsing
and 25 with power pulsing, which are better than the previous results.
The S/N ratio with 6.0 pF feedback capacitance is around 10, which is significantly worse than the cases of 1.2 pF. This indicates an issue in low gain amplification, which is necessary to maximize the dynamic range of SKIROC2/SKIROC2A.

\section{The time measurement with TDC on SKIROC2A}

The arrival time of charges inside the bunch clock (2-5 MHz) can be acquired by
TDC (Time-to-Digital conversion) function in SKIROC2.
The implementation of TDC is improved in the SKIROC2A (which is practically
not usable in SKIROC2).
The TDC in SKIROC2A is implemented with a linearly scanned voltage source
hold by the trigger. The hold voltage is proportional to the arrival time
of the trigger.

\begin{figure}[htbp]
\begin{center}
\includegraphics[width=80mm]{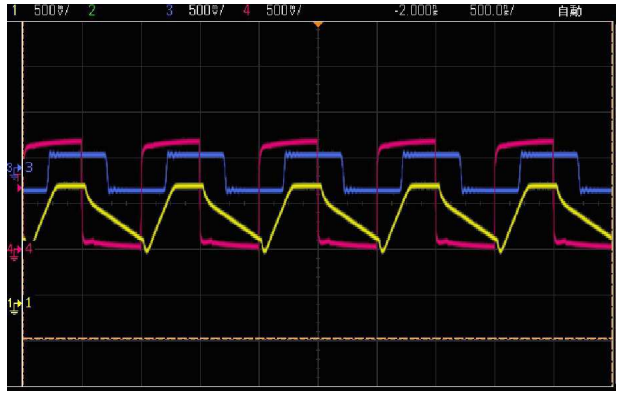}
\caption{A scope view of the test setup. The yellow line shows the TDC ramp voltage with its trigger clock in the pink line. The injected test pulse is shown as the blue line.}
\end{center}
\label{fig:pulse}

\end{figure}

Figure \ref{fig:pulse} shows the relation of clocks and the TDC voltage in our test setup.
The TDC ramp voltage (yellow), synchronized with the TDC clock (pink),
is hold by the trigger issued at the test pulse (blue).
The active time of the ramp voltage is about 200 ns, which corresponds to
the full range of 5 MHz bunch duration.
The provided TDC clock was accidentaly misconfigured to 1 MHz, which causes arbitrary voltage of the ramp voltage after the linear scan, but this expects no significant effects on the measurement at the time of linear scan.

\begin{figure}[htbp]
\begin{center}
\includegraphics[width=60mm]{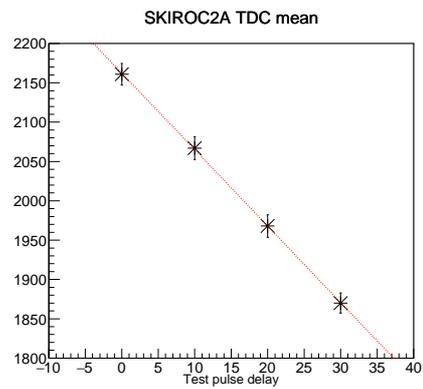}
\caption{ The slope and the resolution of TDC. The horizontal axis is the TDC count, and the vertical axis is the time difference. The error bars show the standard deviation of the TDC distribution with fixed time difference at each point.}
\end{center}
\label{fig:TDC}
\end{figure}

We measured the TDC output with variable time duration between the TDC clock
and the test pulse to obtain the relation between TDC output and the time difference (TDC slope).
We also obtained the time resolution by the distribution of TDC output with fixed time differences.
The results are shown in Fig.~\ref{fig:TDC}.  The TDC slope is around 10 TDC count per ns, and the time resolution is around 1.4 ns with 10 MIP signal injection.

\section{Summary and prospects}

We are studying performance of SKIROC2 and SKIROC2A with BGA testboards.
For the trigger, we obtained the S/N ratio around 13-14 with feedback capacitance of around 1.2 pF.
No significant difference is seen between SKIROC2 and SKIROC2A.
We could get reasonable dynamic range of around 13 global DAC counts on individual trigger threshold adjustment of SKIROC2A.
For the slow shaper, we got the S/N ratio around 25 with feedback capacitance of 1.2 pF and power pulsing. For the lower gain S/N ratio was significantly worse.
The TDC measurement showed that the time resolution is around 1.4 ns with 10 MIP signal injection.

We will continue the study to investigate the issues on the electronics found at the ILD SiW-ECAL prototype, with more study on the power pulsing and effects on the capacitance added for stabilize the power supplies. We will also investigate the additional feature of SKIROC2A, such as the data acquisition with external trigger and the automatic gain selection.

\section*{Acknowledgements}
The Testboard 2 with soldered SKIROC2A was cabled by IN2P3/Omega group
after the intensive tests by them. They also gave a great support and advices to us.
This study was done in close collaboration and discussion with ILD SiW-ECAL group, especially
with the people in LLR and LAL of IN2P3.


\begin{thebibliography}{99}
 \bibitem{ILD} H. Baer et al., The International Linear Collider Technical Design Report - Volume 2: Physics, arXiv:1306.6352.
 \bibitem{pfa} M. A. Thomson, Particle flow calorimetry and the PandoraPFA algorithm, Nucl. Instrum.  Meth. A611 (2009) 25-40
 \bibitem{skiroc2} S. Callier et al., SKIROC2, front end chip designed to readout the Electromagnetic CALorimeter at the ILC, 2011 JINST 6, C12040.
 \bibitem{Hiraitalk} H. Hirai et al., Study of SKIROC2, ASIC for ILD ECAL of ILC,\\
 presentation at CALICE collaboration meeting 2016 at Kyushu University,\\
https://agenda.linearcollider.org/event/6892/contributions/33914/attachments/27940/42290/calice2016.pdf
 \bibitem{Hirairef} H. Hirai, Master thesis at Kyushu University (2016),\\
http://epp.phys.kyushu-u.ac.jp/thesis/2016MasterHirai.pdf (Japanese)
\end{thebibliography}
\end{document}